\documentstyle[pre,twocolumn,aps,epsf,floats]{revtex} 

\begin{document}

\draft

\title{Numerical Determination of the Avalanche Exponents\\ 
of the  Bak-Tang-Wiesenfeld Model}

\author{S. L\"ubeck\cite{SvenEmail} and K.~D. Usadel\cite{UsadelEmail} }
\address{Theoretische Tieftemperaturphysik, 
Gerhard-Mercator-Universit\"at Duisburg,\\ 
Lotharstr. 1, 47048 Duisburg, Germany \\}
\author{\vskip -2\baselineskip\small(Received 17 December 1996)\break}
\author{\parbox{14cm}{\small
        \quad We consider the Bak-Tang-Wiesenfeld sandpile
        model on a two-dimensional square lattice of lattice sizes up
        to $L=4096$.
        A detailed analysis of the probability distribution of 
        the size, area, duration and radius of the avalanches will be given.
        To increase the accuracy of the determination of the avalanche
        exponents we introduce a new method for analyzing the data which reduces 
        the finite-size effects of the measurements.
        The exponents of the avalanche distributions differ 
        slightly from previous measurements and estimates obtained
        from a renormalization group approach.
\hfill\break
\leftline{PACS number: 05.40.+j}}}
\address{\vskip +1.5\baselineskip}

%\begin{abstract}
%We consider the Bak-Tang-Wiesenfeld sandpile
%model on a two-dimensional square lattice of lattice sizes up
%to $L=4096$.
%A detailed analysis of the probability distribution of 
%the size, area, duration and radius of the avalanches will be given.
%To increase the accuracy of the determination of the avalanche
%exponents
%we introduce a new method for analyzing the data which reduces 
%the finite-size effects of the measurements.
%The exponents of the avalanche distributions differ 
%slightly from previous measurements and estimates obtained
%from a renormalization group approach.
%\end{abstract}
%\vspace{-13cm}
%{\tiny accepted for publication in Phys.~Rev.~E}
%\vspace{13cm}

%\pacs{05.40.+j}

\maketitle

\setcounter{page}{4095}
\markright{\rm
  Phys. Rev.~E {\bf 55}, 4095 (1997)
  }
\thispagestyle{myheadings}
\pagestyle{myheadings}

\section{Introduction}

Bak, Tang and Wiesenfeld \cite{BAK} introduced 
the concept of self-organized criticality (SOC)
and realized it with the so-called 'sandpile model' (BTW model).   
The steady state dynamics of the system is characterized by
the probability distributions
for the occurrence of relaxation clusters of a certain  size, area,  
duration, etc .
In the critical steady state these probability distributions
exhibit power-law behavior. 
Using the concept of `Abelian Sandpile Models' \cite{DHAR_2}
it is possible to calculate the static properties of the 
model exactly e.g.~the height probabilities, height 
correlations, number of steady state configurations, 
etc \cite{DHAR_2,MAJUM_1,PRIEZ_1,IVASH_1}.
However, the dynamical properties 
of the model, i.e.~the exponents
of the probability distributions, are not known
exactly. 
Numerical simulations yield different values of the
exponents depending on the considered system size and
the used method of analyzing the data (see for instance 
\cite{MANNA_1,MAJUM_2,CHRIS_2,PRIEZ_2,BENHUR}). 
Recently Pietronero et al.~\cite{PIETRO} introduced a renormalization
scheme which allowed them to estimate the avalanche
exponents.
An improvement of this renormalization scheme was given
by Ivashkevich \cite{IVASH_2} who obtained comparable
results.

We investigate the original Bak, Tang and Wiesenfeld model 
on large lattice sizes ($L\le 4096$)
and measured the probability distributions.
Since the numerical investigations of the BTW model
by Manna \cite{MANNA_1} it is known
that the obtained values of the exponents are affected
by the finite size of the system.
These finite size effects have to be taken into account
in order to get the 'real' exponents.
This has been done by extrapolation ($L\to\infty$) from data 
obtained for different $L$ \cite{MANNA_1}.
We could improve this method and are now able to measure
the exponents of the infinite system directly
thus avoiding any extrapolation.
In this way the accuracy of the obtained exponents
is increased significantly.
We also address the question whether the BTW model and the
related sandpile models of Zhang \cite{ZHANG_1} and
Manna \cite{MANNA_2} belong to the same universality class.
Finally, we discuss the assumption that the avalanche 
propagation  can be described as a random walk.

\section{Model}

We consider the two-dimensional BTW model on a square lattice
of size $L \times L$ in which integer variables $h_{i,j}\ge 0$ represent
local heights.
One perturbes the system by adding particles at a randomly chosen site
$h_{i,j}$ according to
\begin{equation}
h_{i,j} \, \mapsto \, h_{i,j}+1\, , \hspace{0.6cm} \mbox{with random }(i,j).
\label{eq:perturbation}
\end{equation}
A site is called unstable if the corresponding  height $h_{i,j}$ 
exceeds a critical value $h_c$, i.e.~if $h_{i,j} \ge h_c$. 
Without loss of generality, we take $h_c=4$ throughout
this work.
An unstable site relaxes, its value is decreased by four
and the neighboring sites are increased by one unit, i.e.
\begin{equation}
h_{i,j}\;\to\;h_{i,j}\,-\,4
\label{eq:relaxation_1}
\end{equation}
\begin{equation}
h_{i\pm 1,j\pm 1}\;\to\;h_{i\pm 1,j \pm 1}\;+\;1
\label{eq:relaxation_2}
\end{equation}
where the update is done in parallel.
We assume open boundary conditions with heights at the
boundary fixed to zero.

System sizes from $L=64$ to $L=4096$ are investigated.
Starting with a lattice of randomly distributed heights 
$h\in\{0,1,2,3\}$ the system is perturbed according to
Eq.~(\ref{eq:perturbation})
and Dhar's 'burning algorithm' is applied in order to check if the 
system has reached the critical steady state \cite{DHAR_2}.
Then we start the actual measurements.
All measurements are averaged over
at least $10^6$ non-zero avalanches except of the case 
$L=4096$ where only $5 \times 10^5$ measurements have been performed.
We studied four different properties characterizing
an avalanche.  
In the following we use the same notation as Majumdar et al. \cite{MAJUM_2}.
The total number of toppling events is called the size $s$ of
an avalanche. 
The number of distinct toppled lattice sites is denoted by 
$s_d$.
Because a particular lattice site may topple several times the 
number of toppling events exceeds the number of
distinct toppled lattice sites, i.e. $s \ge s_d$.
The duration $t$ of an avalanche is equal to
the number of update sweeps needed until all sites are
stable again.

\begin{figure}
 \begin{minipage}[t]{8.6cm}
 \epsfxsize=8.6cm
 \epsfysize=8.6cm
 \epsffile{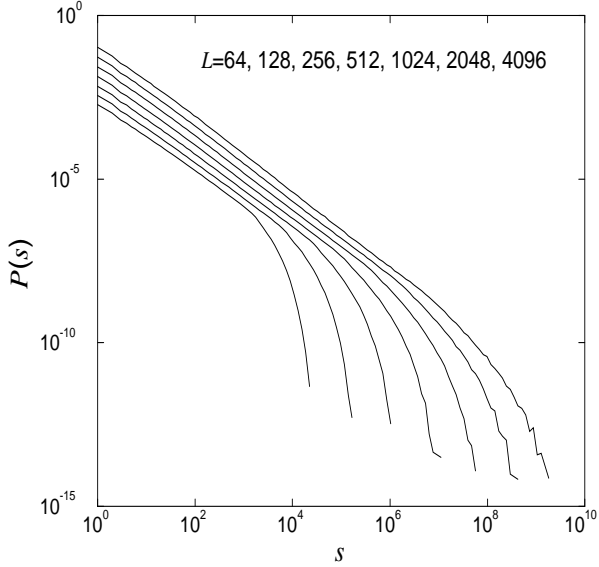} 
 \caption{The probability distribution $P(s)$ for different system sizes.
          The curves for $L<4096$ are shifted in the downward direction.
 \label{distinct_prob}}
 \end{minipage}
\end{figure}

The linear size of an avalanche $r$ is measured via the
radius of gyration of the avalanche cluster.
In the critical steady state the corresponding probability 
distributions should obey power-law behavior 
characterized by exponents $\tau_s$, $\tau_d$,  $\tau_t$ and $\tau_r$
according to
\begin{equation}
P_s(s) \, \sim \, s^{-{\tau_s}},
\label{eq:prob_size}
\end{equation}
\begin{equation}
P_d(s_d) \, \sim \, {s_d}^{-{\tau_d}},
\label{eq:prob_distinct}
\end{equation}
\begin{equation}
P_t(t) \, \sim \, t^{-{\tau_t}},
\label{eq:prob_duration}
\end{equation}
\begin{equation}
P_r(r) \, \sim \, r^{-{\tau_r}}.
\label{eq:prob_radius}
\end{equation}

\section{Simulations and Results}

\begin{figure}
 \begin{minipage}[t]{8.6cm}
 \epsfxsize=8.6cm
 \epsfysize=8.6cm
 \epsffile{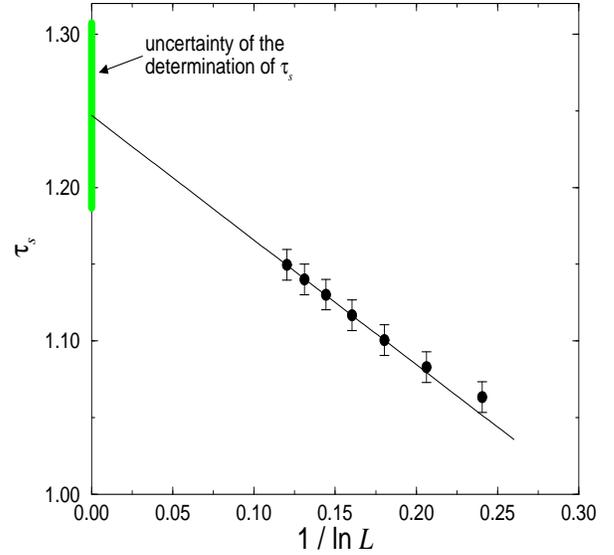} 
 \caption{Determination of the exponent $\tau$ using 
          the extrapolation (Eq.~\protect\ref{eq:tau_L}).
 \label{tau_reg}} 
 \end{minipage}
\end{figure}

Fig.~\ref{distinct_prob} displays the obtained results
for the distribution $P_s(s)$  for different system sizes.
A power-law fit to the straight portion of these curves
yields the exponents $\tau_s(L)$.
Fig.~\ref{tau_reg} shows a plot of the exponents $\tau_s(L)$ 
vs.~$1/\ln{L}$.
It is seen that
for $L \ge 128$ the exponents obey the finite-size behavior
\begin{equation}
\tau_s(L) = \tau_{s,\infty} \, - \, \frac{const}{\ln{L}}
\label{eq:tau_L}
\end{equation}
as suggested already by Manna \cite{MANNA_1}.
The extrapolation to $L\to\infty$ yields the value
of the exponent $\tau_{s,\infty}=1.247$.
The probability distributions $P_d(s_d)$, $P_t(t)$ 
and $P_r(r)$
are analyzed in the same way with the result 
$\tau_{d,\infty}=1.258$, $\tau_{t,\infty}=1.405$ and 
$\tau_{r,\infty}=1.588$, respectively.
All exponents are slightly larger than those
obtained from earlier simulations by Manna
who considered smaller system
sizes and had less statistics\cite{MANNA_1}.

However, these values of the exponents are not very accurate.
Namely, a crucial point in this analysis is the extension
of the fit region in each distribution $P_s(s,L)$.
Changing it slightly different exponents are obtained.
This uncertainty in the determination
of the exponents $\tau(L)$ can be estimated to be
at least of the order of $\pm0.01$.
Taking then the propagation of these errors into account we
can estimate the uncertainty in the determination of the 
extrapolated value $\tau_{\infty}$ to be of the order 
of $\pm 0.06$ which is mainly due to the large distance of
the measured values from the vertical axis (see Fig.~\ref{tau_reg}).
Thus it is in principle not possible to obtain the exponents 
of the BTW model with high accuracy by a simple 
extrapolation of the exponents via Eq.~(\ref{eq:tau_L}).

However, it is possible to improve the determination of 
the exponents not by using Eq.~(\ref{eq:tau_L}) for an
extrapolation but for a direct determination of $\tau_{\infty}$.
Consider for this purpose two probability 
distributions $P(s,L_1)$ and $P(s,L_2)$ corresponding
to different system sizes with $L_1>L_2$. 
If Eq.~(\ref{eq:tau_L}) 
describes the finite-size behavior of the exponents $\tau_s$
correctly the probability distribution (Eq.~\ref{eq:prob_size}) 
for a given system size $L$ behaves as
\begin{equation}
P(s,L) \; \sim \; s^{-\tau_{s,\infty}} \; s^{\frac{const}{\ln{L}}}. 
\label{eq:prob_L}
\end{equation}
Thus, the exponent $\tau_{s,\infty}$ can be determined directly by a
power-law fit of the function $H(s,L_1,L_2)$
which is defined as
\begin{equation}
H(s,L_1,L_2) \; = \;
\frac{P(s,L_1)^{\ln{L_1}}}{P(s,L_2)^{\ln{L_2}}}
\; \sim \;
s^{-\tau_{s,\infty} \, (\ln{L_1}-\ln{L_2}) }.
\label{eq:help_fkt}
\end{equation}

\begin{figure}
 \begin{minipage}[t]{8.6cm}
 \epsfxsize=8.6cm
 \epsfysize=8.6cm
 \epsffile{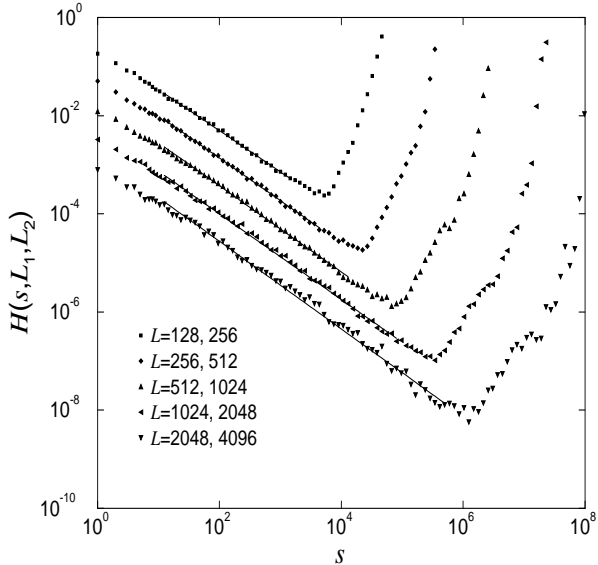} 
 \caption{ The function $H(s,L_1,L_2)$ for different pairs $L_1$ and $L_2$.
           The curves are shifted with increasing system sizes in the 
           downward direction. The solid lines correspond to a power-law fit. 
	   The obtained values of the exponent $\tau_s$ are listed in 
           Table~{\protect\ref{TABLE_TAU}}.
 \label{help_funktion}}
 \end{minipage}
\end{figure}

In Fig.~\ref{help_funktion} $H(s,L_1,L_2)$ is plotted for various
system sizes $L_1$ and $L_2$.
A nice property of this function is that 
in contrast to the probability distribution the cut off of the
power-law behavior at large values of $s$ is now very abrupt.
We apply this analysis to all four distributions and the
resulting exponents are listed in Table~\ref{TABLE_TAU}.
The values of the exponents $\tau_{s,\infty}$, 
$\tau_{t,\infty}$ and $\tau_{r,\infty}$ (except of the case $L_2=128$, $L_1=256$) 
fluctuate around their mean values given by $\tau_{s,\infty}=1.293\pm0.009$, 
$\tau_{t,\infty}=1.480\pm0.011$ and $\tau_{r,\infty}=1.665\pm0.013$.
Only the exponent $\tau_{d,\infty}$ displays a significant
$L$-dependence.
A  possible origin of this $L$-dependence is that Eq.~(\ref{eq:tau_L})
does not describe correctly the finite size behavior of $\tau_d$
and that one has to add corrections to it.
However, the data suggest that this additional $L$-dependence vanishes 
for large system sizes
and therefore the exponent saturates in the vicinity 
of $\tau_{d,\infty} \approx 1.33$. 
Note that the mentioned error-bars describe only the 
statistical error.
Due to the systematic errors the real error-bars are slightly
larger.

\section{Discussion}

Despite of their different toppling rules it is supposed that the BTW 
model, Zhang's model \cite{ZHANG_1} and Manna's Two-State 
model \cite{MANNA_2} belong to the same universality 
class, i.e.~they should be characterized by the same exponents.
Pietronero and co-workers \cite{PIETRO} addressed this question
by a renormalization group approach and found that the
BTW model and Manna's Two-State model belong to the 
same universality class.
Different results were obtained by Ben-Hur and Biham \cite{BENHUR}
who found different values for the two models.

\vspace{0.2cm}
\begin{table}
\caption{Values of the exponents $\tau_s$, $\tau_d$, $\tau_t$ 
and $\tau_r$ for different pairs of system sizes $L$.}
\label{TABLE_TAU}
\begin{tabular}{lcccc}
$L_1,L_2$ & $\tau_{s,\infty}$  & $\tau_{d,\infty}$  & $\tau_{t,\infty}$ & $\tau_{r,\infty}$ \\  
\tableline
$128, 256$   & 1.293  & 1.253 & 1.486 & 1.183 \\
$256, 512$   & 1.281  & 1.287 & 1.464 & 1.665 \\
$512, 1024$  & 1.305  & 1.328 & 1.487 & 1.648 \\ 
$1024, 2048$ & 1.286  & 1.330 & 1.479 & 1.684 \\
$2048, 4096$ & 1.298  & 1.331 & 1.483 & 1.661 \\
\end{tabular}
\end{table}

\begin{table}[b]
\caption{Values of the exponents $\tau_s$, $\tau_d$, $\tau_t$ 
and $\tau_r$ for the BTW model, Zhang's model 
and Manna's Two-State model.
Due to a finite curvature of the probability distribution 
the duration exponent $\tau_t$ of the Zhang model 
can not be determined in the usual way \protect\cite{LUEB_3}.}
\label{TABLE_MODEL}
\begin{tabular}{lcccc}
Model& $\tau_{s}$  & $\tau_{d}$  & $\tau_{t}$ & $\tau_{r}$ \\  
\tableline
BTW   & 1.293  & 1.330 & 1.480 & 1.665 \\
Zhang & 1.282  & 1.338 &       & 1.682  \\
Manna & 1.275  & 1.373 & 1.493 & 1.743  \\ 
\end{tabular}
\end{table}

In Table~\ref{TABLE_MODEL} we compare our results with the 
exponents of the Zhang and the Two-State model obtained 
from recent investigations on comparable lattice 
sites \cite{LUEB_3}.
Within the error-bars the BTW and the Zhang model displays the
same exponents.
The differences of the exponents $\tau_d$ and $\tau_r$ of the
BTW and Manna's model can not be explained by the error-bars
and thus we conclude that both models
do not belong to the same universality class.
But it is remarkable  that both models display 
nearly the same duration exponent $\tau_t$ and especially that
$\tau_t \approx \frac{3}{2}$.
We assume that the value $\tau_t=\frac{3}{2}$ is a 
common feature of many sandpile models
caused by an analogy of the avalanche propagation and a
random walk, which we will discuss now.

The number of critical sites $n(t)$ at a given update (time) 
step $t$ can be considered as a random walker.
Starting with $n(t=0)=1$ the avalanche performs a random
walk $n(t=0)\to n(t=1) \to n(t=2) \to ...$ with
the transition probabilities $p(n,n')$.
The avalanche ceases to exist if the random walk returns to the
origin ($n=0$).
In the simplest case the transition probabilities are
homogeneous $p(n,n')=p(n-n')$, 
symmetric $p(\Delta n)=p(-\Delta n)$ and
the random numbers $\Delta n$ are uncorrelated.
Then the  avalanche probability distribution $P_t(t)$ is given by the
probability $P_{\mbox{\tiny first return}}(t)$ that a random walker 
with initial value $n(t=0)=1$, with certain transition probabilities for
increasing, decreasing and maintaining $n$, returns
for the first time to its starting point in $t$ steps, 
which scales as \cite{GRIMMETT}
\begin{equation}
P_{\mbox{\tiny first return}}(t)\sim t^{-\frac{3}{2}}.
\label{eq:rand_walk}
\end{equation}
Certain sandpile models are solved by an exactly mapping 
of the avalanche propagation onto a simple random 
walk \cite{DHAR_1,FLYVBJERG_2}.

\begin{figure}
 \begin{minipage}[t]{8.6cm}
 \epsfxsize=8.6cm
 \epsfysize=8.6cm
 \epsffile{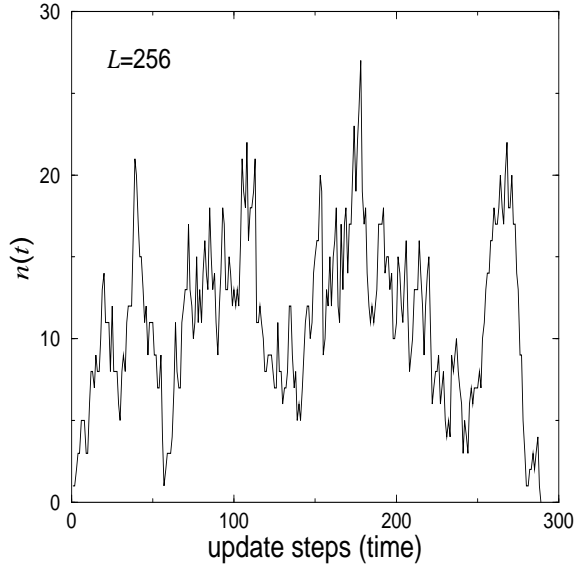} 
 \caption{The avalanche propagation as a random walk. The number
	  of critical sites $n(t)$ is plotted against the update
	  (time) steps for a certain avalanche of duration $t=289$.
	  Starting from $n(t=0)=1$ the avalanche stops if the 
	  random walker returns to the origin for the first time.
\label{rand_walk_1}} 
 \end{minipage}
\end{figure}

\begin{figure}
 \begin{minipage}[t]{8.6cm}
 \epsfxsize=8.6cm
 \epsfysize=8.6cm
 \epsffile{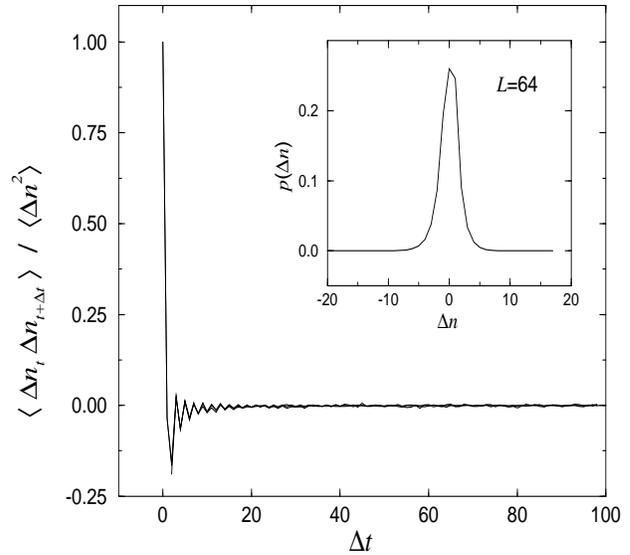} 
 \caption{The probability distribution $p(\Delta n)$ (see inset)
	and the correlations $C(\Delta t)$ for different system sizes
	($L=64, 128, 256$). 
 \label{rand_walk_2}}  
 \end{minipage}
\end{figure}

In Fig.~\ref{rand_walk_1} we present the number of critical 
sites vs. update steps of a certain avalanche of the BTW model.
The probability distribution $p(\Delta n)$ and the 
corresponding correlation function
\begin{equation}
C(\Delta t) \; = \;
\frac{\langle \, \Delta n(t) \, \Delta n(t+\Delta t)\, \rangle}
{\langle \, \Delta n^2\, \rangle } 
\label{eq:rw_corr}
\end{equation}
are shown in Fig.~\ref{rand_walk_2}.
The probability distribution $p(\Delta n)$ has to be symmetric
in order to make sure that the random walk is recurrent, 
i.e.~the probability that it ever returns to the origin 
is one \cite{GRIMMETT}.
The distribution displays asymmetries only for finite system 
sizes.
A detailed analysis (not shown) yields that the third central moment 
of the distribution $p(\Delta n)$ tends to zero with 
diverging system size $L$ indicating 
that $p_{L\to\infty}(\Delta n)$ is symmetric.

The correlation function $C(\Delta t)$ is sharply peaked at 
$\Delta t=0$ but there are small oscillations for small
values of $\Delta t$.
Therefore, the second requirement for Eq.~(\ref{eq:rand_walk})
to be valid, uncorrelated steps $\Delta n$, is only
fulfilled approximately.
This oscillating behavior is caused by the used parallel 
update process.
Since toppling occurs at a given time step in one 
sublattice only the update algorithm
switches in sequently time steps between the two sublattices.
The alternating correlation function indicates that the 
correlations within one sublattice differs from the correlation
between the two sublattices.
Thus, compared to the exact solved sandpile models 
\cite{DHAR_1,FLYVBJERG_2} where the correlation functions
are simply given by a $\delta$-function the correlations
of the BTW model are more complicated.
But since these oscillations at small $\Delta t$
have small amplitudes we suggest that the 
avalanche propagation may be described as a random walk 
and that the exponent of the duration is $\tau_t=\frac{3}{2}$.

Scaling relations for the exponents $\tau_s, \tau_d, \tau_t$ and
$\tau_r$ can be obtained if one assumes that the 
size, area, duration and radius scale as a 
power of each other, for instance 
\begin{equation}
t \, \sim \, r^{\gamma_{tr}}
\label{eq:gam_tr}
\end{equation}
for the duration $t$ of an avalanche and its radius $r$.
The relation $P_t(t) \mbox{d}t=P_r(r) \mbox{d}r$ for the
corresponding distribution functions leads to
the scaling relation
\begin{equation}
{\gamma_{tr}}\;=\;\frac{\tau_r-1}{\tau_t-1}.
\label{eq:gam_tau_tr}
\end{equation}
The exponents $\gamma_{dr}$, $\gamma_{rs}$,
$\gamma_{sd}$ etc are defined in the same way.
The exponent $\gamma_{tr}$ is usually identified
with the dynamical exponent $z$ and using a 
momentum-space analysis of the corresponding
Langevin equations D\'{\i}az-Guilera showed that
the dynamical exponent of the BTW and Zhang's model 
is given by $z=(d+2)/3$ \cite{DIAZ_1}.
On the other hand one concludes from the compactness
of the avalanche clusters that $\gamma_{dr}=2$.
Thus one gets two scaling relations for
the exponents $\tau_d$, $\tau_r$ and $\tau_t$
and using the result that $\tau_t=\frac{3}{2}$
the exponents of the probability distribution
of the radius and the area are given  
by $\tau_r=\frac{5}{3}$ and $\tau_d=\frac{4}{3}$.
These values are in good agreement with our numerical
results and we would suggest that they are
the exact exponents of the BTW model.

Majumdar and Dhar~\cite{MAJUM_2} assumed that the 
size and the area of an avalanche fulfill the
relation
\begin{equation}
s \; \sim \; s_d\,n_c 
\label{eq:Majumdar_relation}
\end{equation}
where $n_c$ is the number of topplings at the
site initiating the avalanche. 
If this equation holds the exponents $\tau_s$ 
and $\tau_d$ have to fulfill the relation
$\tau_s =2-1/ \tau_d$.
Using  $\tau_d=\frac{4}{3}$ from above we obtain
$\tau_s=\frac{5}{4}$ which is well outside the error-bars
of our numerical result $\tau_s=1.293$.
Thus we conclude that the assumed 
relation Eq.~(\ref{eq:Majumdar_relation}) does not
describe the real scaling behavior. 

\begin{figure}
 \begin{minipage}[t]{8.6cm}
 \epsfxsize=8.6cm
 \epsfysize=8.6cm
 \epsffile{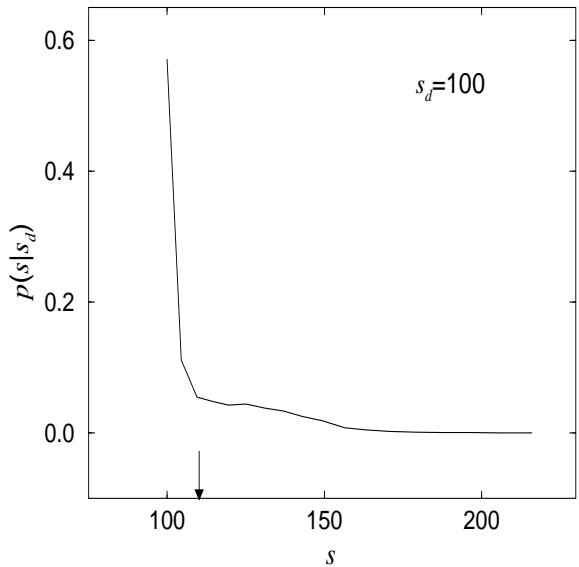} 
 \caption{The conditional probability distribution $p(s|s_d)$.
          The arrow marks the corresponding expectation value.
 \label{cond_prob}} 
 \end{minipage}
\end{figure}

Due to a lack of a scaling relation which connects $\tau_s$
with the other exponents the exact value of the exponent $\tau_s$
is still unknown.
Even a numerical determination of the exponent $\gamma_{sd}$
yields useless results.
The relation which 
defines $\gamma_{sd}$ implies the assumption that 
the conditional probability distribution
$p(s|s_d)$ is strongly peaked so that the expectation 
value $E(s|s_d)$ scales with the area $s_d$.
Measurements of the conditional probabilities show
that this is not the case for $p(s|s_d)$ (see Fig.~\ref{cond_prob}).
The distribution displays an asymmetric shape 
which violates the above assumptions.

A similar analysis of Manna's Two-State model yielded that
the dynamical exponent is given by $z\approx \frac{3}{2}$ resulting
in $\tau_r=\frac{7}{4}$, $\tau_d=\frac{11}{8}$ \cite{LUEB_3}.
The BTW model and the Two-State model belongs to
different universality classes.

\section{Conclusions}

We studied numerically the dynamical properties of the BTW model on a 
two-dimensional square lattice and measured for large
system sizes ($L\le 4096$) the avalanche 
probability distributions.
We introduced a new analysis to minimize the finite-size effects
and determined the avalanche exponents with an improved accuracy.
Our numerical results are consistent with the values
$\tau_t=\frac{3}{2}$, $\tau_r=\frac{5}{3}$ and $\tau_d=\frac{4}{3}$
which we consider to be the exact exponents of the BTW model.
We discussed the possibility that these values are caused
by an analogy of the avalanche propagation and a
random walk process. 
Further work has to be done to check this assumption. 
Recently, Ivashkevich \cite{IVASH_2} improved the renormalization group approach 
for sandpile models proposed by Pietronero et al.~\cite{PIETRO}.
Both calculation yields the exponent $\tau_d\approx 1.25$ 
significantly smaller than our numerical estimates.

\acknowledgments
This work was supported by the
Deutsche Forschungsgemeinschaft through
Sonderforschungsbereich 166, Germany.
We would like to thank D.~Ktitarev for helpful discussions.

\end{document}